\documentclass[osajnl,twocolumn,showpacs,superscriptaddress,10pt]{revtex4-1} %% use 11pt for Applied Optics
\usepackage{textcomp,amsmath,amssymb,graphicx}
\newcommand{\ic}{\affiliation{Blackett Laboratory, Imperial College London, Prince Consort Road, London SW7 2AZ, UK}}
\newcommand{\loa}{\affiliation{Laboratoire d'Optique Appliqu\'ee, Ecole Nationale Sup\'erieure de Techniques Avanc\'ees--ParisTech, Ecole Polytechnique-CNRS, 91761 Palaiseau Cedex, France}}

\begin{document}

\title{Carrier-envelope phase stability of hollow-fibers used for high-energy, few-cycle pulse generation}

\author{William A. Okell}\email{Corresponding author: william.okell09@imperial.ac.uk}
\author{Tobias Witting}
\author{Davide Fabris}\ic
\author{Dane Austin}\ic
\author{Ma\"imouna Bocoum}\loa
\author{\\Felix Frank}\ic
\author{Aur\'elien Ricci}
\author{Aur\'elie Jullien}\loa
\author{Daniel Walke}\ic
\author{Jonathan P. Marangos}\ic
\author{\\Rodrigo Lopez-Martens}\loa
\author{John W. G. Tisch}\ic

\begin{abstract}
We investigated the carrier-envelope phase (CEP) stability of a hollow-fiber setup used for high-energy, few-cycle pulse generation.  Saturation of the output pulse energy is observed at $0.6\,$mJ for a $260\,$\textmu{}m inner-diameter, $1\,$m long fiber, statically filled with neon, with the pressure adjusted to achieve an output spectrum capable of supporting sub-$4\,$fs pulses. The maximum output pulse energy can be increased to $0.8\,$mJ by using either differential pumping, or circularly polarized input pulses.  We observe the onset of an ionization-induced CEP instability, which does not increase beyond an input pulse energy of $1.25\,$mJ due to losses in the fiber caused by ionization. There is no significant difference in the CEP stability with differential pumping compared to static-fill, demonstrating that gas flow in differentially pumped fibers does not degrade the CEP stabilization.
\end{abstract}

(320.0320) Ultrafast optics; (320.5520) Pulse compression; (320.7090) Ultrafast lasers; (320.7100) Ultrafast measurements; (320.7110) Ultrafast nonlinear optics; (320.7140) Ultrafast processes in fibers.% REPLACE WITH CORRECT OCIS CODES FOR YOUR ARTICLE
                          % NOTE: \ocis{} IS ALIASED TO \pacs{} BUT MUST
                          % FORMAT THE TERMS CORRECTLY FOR EACH JOURNAL

\maketitle %% required

High-energy, few-cycle pulses can be generated with millijoule-level, $20$--$30\,$fs pulses from a Ti:sapphire chirped pulse amplifier (CPA) using hollow-fiber pulse compression \cite{nisoli_generation_1996}. Pulses as short as $3.5\,\mathrm{fs}$, with $0.2\,$mJ energy \cite{witting_characterization_2011}, and pulses with up to $5\,\mathrm{mJ}$ of energy in somewhat longer pulses of $5\,$fs duration \cite{bohman_generation_2010}, have been generated using this technique. For a typical fiber inner-diameter of $250\,$\textmu m, there is a trade-off between spectral broadening and energy transmission because ionization and self-focusing become increasingly important issues when the input pulse energy exceeds $\sim\!1\,$mJ.  To combine a high throughput with a large broadening factor, differential pumping (DP) \cite{suda_generation_2005,robinson_the_2006} or circularly polarized (CP) input pulses \cite{chen_generation_2009}, can be used to mitigate the impact of these unwanted nonlinear effects. The continuing development of hollow-fiber technology towards single-cycle pulses with $>\!1\,$mJ energy is particularly important for relativistic laser-matter studies \cite{borot_attosecond_2012}, and for the generation of more intense attosecond pulses. While larger diameter fibers are also a promising route to higher energy few-cycle pulses \cite{bohman_generation_2010,nagy_optimal_2011,schweinberger_waveform_2012}, the looser focusing conditions demand a significantly increased laboratory space for the apparatus in order to minimize nonlinear effects in the entrance and exit windows.

Although stabilization of the carrier-envelope phase (CEP) is crucial to many experiments requiring few-cycle pulses with higher energies, the effect that energy scaling of hollow-fiber pulse compression has on the CEP stability of the output pulses is yet to be investigated systematically. Through self-phase modulation (SPM), the main spectral broadening process occurring in the fiber, pulse energy fluctuations induce a CEP instability in the output pulses \cite{wang_coupling_2009}. At high input pulse energies, ionization within the fiber becomes significant and will also contribute to the CEP fluctuations. Furthermore, the possibility of an additional instability caused by gas flow in differentially pumped fibers has not yet been explored. 

In this Letter, we examine the energy scaling of the hollow-fiber compression technique in three commonly used modes of operation: statically-filled using linearly polarized pulses (SFLP); differentially pumped using linearly polarized pulses (DPLP); and statically-filled using CP pulses (SFCP).  We also investigate the CEP stability of the fiber as a function of the input energy.

In our experiment, $28\,$fs pulses with up to $2.5\,$mJ energy at a 1\,kHz repetition rate are generated by a commercial CPA system (Femtolasers GmbH, Femtopower HE CEP). The CPA is CEP stabilized using a photonic crystal fiber based \it f\rm-to-2\it f \rm interferometer for fast-loop locking. The amplified pulses were delivered to a $260\,$\textmu m inner-diameter, $1\,$m long fiber filled with neon. The experimental setup is shown in Fig.\,1. To statically fill the fiber, both the entrance and exit tubes are filled with neon. Differential pumping is achieved by evacuating the entrance tube, which is typically held at $\sim10^{-1}\,$mbar when the exit tube is filled with $3\,$bar of neon.  An $f=1.5\,\mathrm{m}$ focusing mirror couples the beam into the fiber, and the output beam is recollimated using an $f=87.5\,$cm mirror. The spectrally broadened pulses are compressed using 10 reflections from double-angle technology chirped-mirrors (UltraFast Innovations GmbH), and through fine-tuning of the group delay dispersion (GDD) with fused silica wedges. The pulse can be converted to circular polarization for propagation through the fiber by introducing a quarter-wave plate into the beam.   A broadband achromatic quarter-wave plate (Femtolasers GmbH, OA229) converts the pulses back to linear polarization after the fiber.
\begin{figure}[htbp]
\label{setup}
\centerline{\includegraphics[width=\columnwidth]{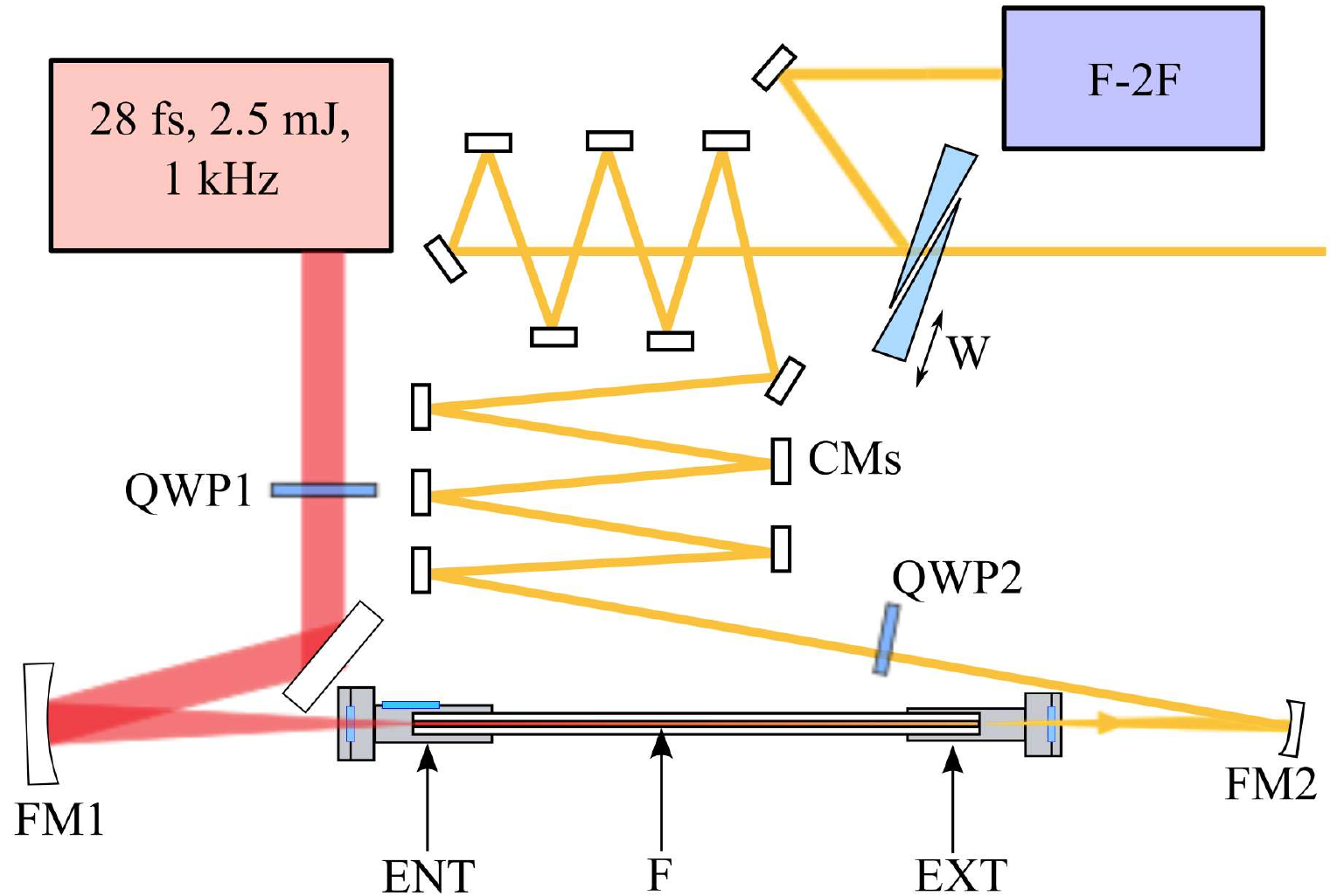}}
\caption{Experimental setup. QWP1,2: quarter-wave plates; FM1,2: focusing and recollimating mirrors, respectively; F: hollow-fiber; ENT, EXT: entrance and exit tubes, respectively; CMs: chirped mirrors; W: fused silica wedges; F-2F: slow-loop \it f\rm-to-2\it f \rm interferometer.}
\end{figure}

The beam reflected from the front-face of the wedges enters an \it f\rm-to-2\it f \rm interferometer, which is used to measure the CEP stability after the fiber. Since the spectrum from the hollow-fiber spans more than an octave, no additional spectral broadening is required within the interferometer.  The pulses are focused into a beta-barium borate (BBO) crystal to double the long wavelength part of the spectrum. A polarizing beamsplitter cube enables interference of the overlapping regions of the second harmonic and fundamental spectra around $520\,$nm. Feedback is achieved by applying a DC-offset to the fast-loop locking electronics.

\begin{figure}[htbp]
\label{pscan}
\centerline{\includegraphics[width=\columnwidth]{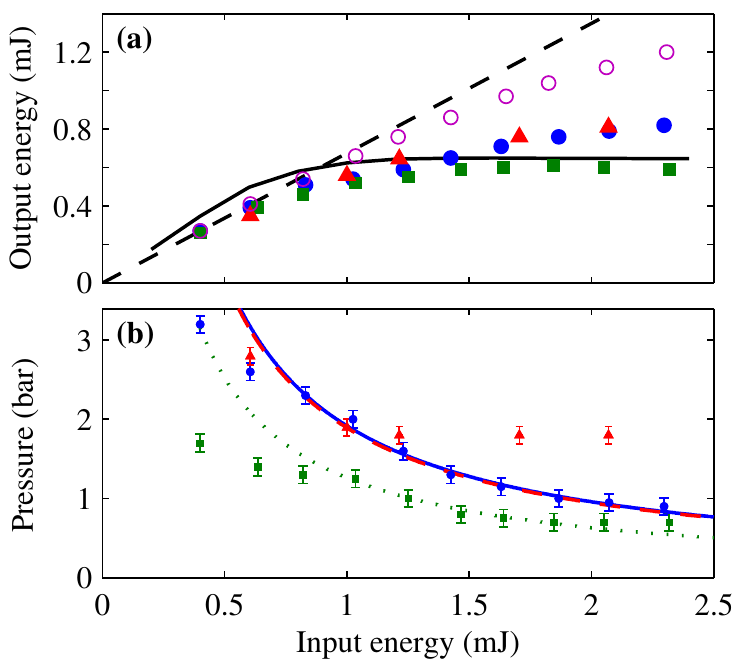}} %20130530_pscan
\caption{(a) Output pulse energy with the fiber evacuated with LP (purple open circles), and for an output spectrum with a sub-$4\,$fs Fourier transform limit for SFLP (green squares), DPLP (blue circles), and SFCP (red triangles).  The black dashed line represents $68\,\%$ transmission. The black solid line is the transmission predicted by our model. (b) Corresponding experimental neon pressure in the exit tube (data points), and theoretical pressure for SFLP (dotted green curve), DPLP (solid blue curve), and SFCP (dashed red curve).}
\end{figure}
The input pulse energy was varied using a half-wave plate and polarizer before the compressor in the CPA, while the neon pressure was adjusted to maintain a spectrum with a sub-$4\,$fs Fourier transform limit (FTL) at the output of the fiber. The output pulse energy is shown as a function of the input pulse energy in Figure $2\,$(a). Below an input pulse energy of $0.6\,$mJ, the energy transmission for all configurations is close to the vacuum transmission at low energy ($68\,\%$). At input pulse energies above $1.5\,$mJ, the output energy for SFLP saturates at $0.6\,$mJ. Both SFCP and DPLP permit significantly higher output pulse energies of up to $0.8\,$mJ, but also show deviation from the vacuum transmission at high input pulse energies, and appear to be approaching saturation. Figure $2\,$(b) shows the experimental and theoretical neon pressure in the exit tube as a function of the input pulse energy. The theoretical curves were calculated using the broadening factor defined in \cite{vozzi_optimal_2005}, which leads to a gas pressure given by
\begin{equation}
p = \frac{cA_{\mathrm{eff}}}{2\kappa_{2}\omega_{0}P_{0}L_{\mathrm{eff}}}\left[3\sqrt{3}\left(F^{2} - 1\right)\right]^{1/2},
\end{equation}
where $c$ is the speed of light in a vacuum, $A_{\mathrm{eff}}\approx0.48\pi a^{2}$ is the effective area of the fiber of inner-radius $a$, $\kappa_{2} = 7.4\times10^{-25}\,$m$^{2}\,$W$^{-1}\,$bar$^{-1}\,$ is the neon nonlinear coefficient per unit pressure \cite{nisoli_toward_1998}, $\omega_{0}$ is the laser central frequency, $F = \Delta\omega/\Delta\omega_{0}$ is the broadening factor, $\Delta\omega_{0}$ and $\Delta\omega$ are the initial and final bandwidths, respectively, and $P_{0}$ is the laser peak power. The effective fiber lengths, $L_{\mathrm{eff}}$, for SFLP, SFCP and DPLP are
\begin{align}
 L_{\mathrm{eff}}^{\mathrm{SFLP}}&=\frac{1-\mathrm{e}^{-\alpha L}}{\alpha},\;\;\nonumber
L_{\mathrm{eff}}^{\mathrm{SFCP}}=\frac{2}{3}\left(\frac{1-\mathrm{e}^{-\alpha L}}{\alpha}\right),\\
L_{\mathrm{eff}}^{\mathrm{DPLP}}&\approx\frac{2}{3}\left(\frac{1-\mathrm{e}^{-1.21\alpha L}}{1.21\alpha}\right),
\end{align}
respectively, where $L$ is the fiber length and $\alpha$ is the mode attentuation constant defined in \cite{vozzi_optimal_2005}. For our experimental parameters, $L_{\mathrm{eff}}^{\mathrm{SFLP}}=0.94\,$m, $L_{\mathrm{eff}}^{\mathrm{SFCP}}=0.63\,$m, and $L_{\mathrm{eff}}^{\mathrm{DPLP}}=0.62\,$m.

\begin{figure}[htbp]
\label{CEP}
\centerline{\includegraphics[width=\columnwidth]{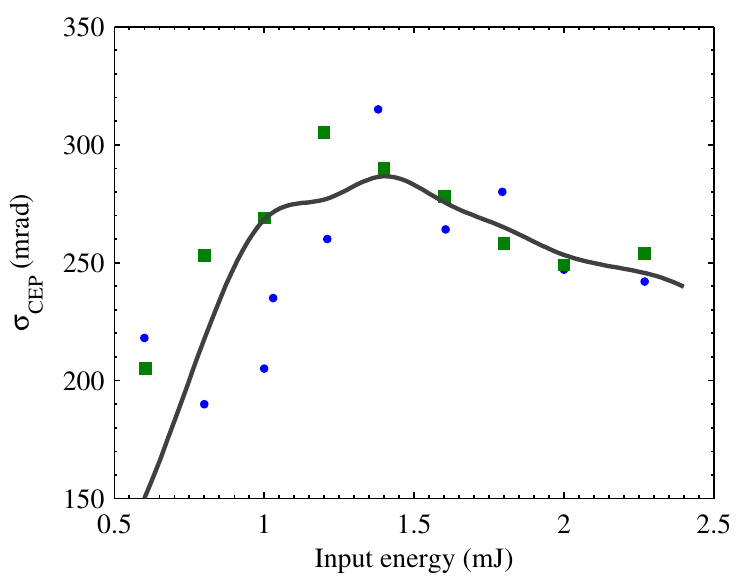}} %CEP_plot
\caption{Experimental standard deviation of CEP fluctuations, $\sigma_{\mathrm{CEP}}$, measured after the fiber using DPLP (blue circles) and SFLP (green squares), and theoretical CEP fluctuations predicted by propagation simulations (black line).}
\end{figure}
%!!!!!TODO add clever sentence linking to power saturation TODO!!!!!
The saturation of the output energy is an indication that physical processes other than SPM start to play a significant role. Therefore we investigate the CEP stability as a function of input energy. A plasma with a free electron density $\rho_{e}$ will induce a CEP shift of $\varphi\approx e^{2}\rho_{e}z/m_{e}\epsilon_{0}c\omega_{0}$ over a distance $z$, where $\omega_{0}$ is the laser central frequency, $c$ is the speed of light, $\epsilon_{0}$ is the permittivity of free-space, and $e$ and $m_{e}$ are the electronic charge and mass, respectively. Modern CPA laser systems typically have output pulse energy fluctuations of around $1\,\%$, which will cause fluctuations in the plasma density in the fiber. The CEP fluctuations induced as a consequence are given by
\begin{equation}
\delta\varphi = \frac{e^{2}z}{m_{e}\epsilon_{0}\omega_{0}c}
\frac{\partial\rho_{e}(I_{0},p)}{\partial I_{0}}\delta I_{0},
\end{equation}
where $I_{0}$ is the peak intensity of the laser pulse, $\delta I_{0}$ is a small change in the pulse peak intensity, and $p$ is the gas pressure. Pulse energy fluctuations can also induce a CEP instability associated with SPM, through the mechanism described in \cite{wang_coupling_2009}. In order to model the CEP fluctuations induced by the fiber, we have simulated our experiments using a coupled-mode, split-step technique incorporating modal dispersion and loss, the Kerr effect including self-steepening, and ionization. The ionization rate was calculated using the Ammosov, Delone and Krainov theory \cite{ammosov_tunnel_1986}. The initial conditions were the experimentally measured laser temporal profile, and a Gaussian spatial profile with an optimal $1/e^{2}$ diameter of $0.64\!\times\!260\,$\textmu m. The gas pressure was varied with input energy in order to maintain a constant broadening  factor, as prescribed by Equation 1. The input pulses were assumed to have $1\,\%$ energy fluctuations and $\sim150\,$mrad CEP fluctuations, which is the typical performance of our laser system. Our simulations were performed for DPLP, where in-coupling effects due to ionization and self-focusing of the beam  can be neglected due to the low gas pressure ($<10^{-1}\,$mbar).

Figure $3\,$ shows the standard deviation of the CEP measured after the fiber using DPLP and SFLP. The overall trend for both DPLP and SFLP is a degradation in the CEP stability as the input pulse energy is increased, suggesting the onset of an ionization-induced CEP instability. The results of the simulation are also shown in Figure $3\,$, and show excellent agreement with the experimental data. The simulation confirms that the decrease in the CEP stability as the input energy is increased from $0.5\,$mJ to $1.25\,$mJ is a consequence of ionization in the fiber. Above $1.25\,$mJ, further degradation in the CEP stability is prevented by energy losses caused by ionization. Without direct ionization losses (which include the acceleration of ionized electrons), this roll-off remains present, and can therefore be attributed to ionization defocusing causing energy losses by coupling energy into higher order modes of the fiber.
%Below $0.5\,$mJ, simulations performed with no ionization effects reveal that the CEP instabilities induced by the fiber are dominated by fluctuations in the SPM process. We are unable to perform simulations with zero ionization above $0.6\,$mJ because an unphysical shock front develops at the trailing edge of the pulse, due to low dispersion caused by low gas pressures.

The experimental results show no significant difference in the CEP stability using DPLP compared to SFLP, demonstrating for the first time that gas flow in differentially pumped fibers does not degrade the CEP stability. However, in a DP fiber, the highest gas density is at the fiber exit. At the exit, mode attenuation will have decreased the peak intensity of the pulse. Since the ionization-induced CEP instability is linear with pressure, but extremely nonlinear with intensity, one might expect a CEP stability improvement when using DPLP compared to SFLP, in parameter regimes where the ionization-induced CEP instabilities are not prevented by propagation losses. %The CEP stability of SFCP will be explored in a separate study.

\begin{figure}[htbp]
\label{CEP_longterm}
\centerline{\includegraphics[width=\columnwidth]{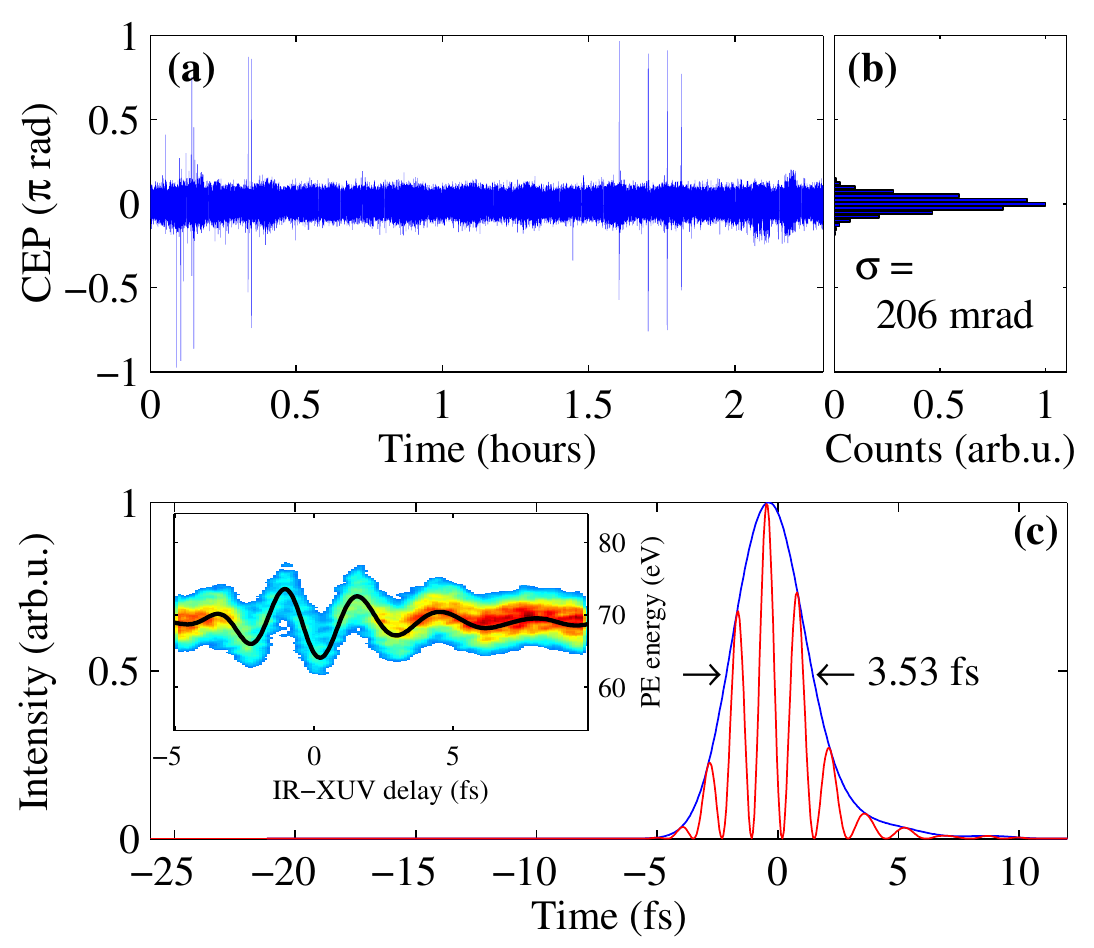}} %CEP_and_It_withCRABinset
\caption{CEP stability and temporal characterization of pulses generated using DPLP. (a) CEP data; (b) histogram of CEP data; (c) full temporal characterization of $0.4\,$mJ compressed pulses, showing the square of the retrieved electric field (red), and the intensity envelope (blue) with a FWHM duration of $3.5\,$fs. The experimental photoelectron (PE) kinetic energy spectrum as a function of the time delay between the IR and XUV pulses (FROG-CRAB trace) is shown in the inset. The retrieved vector potential of the IR pulse is also shown (black line).}
\end{figure}
For CEP sensitive experiments in day-to-day operation we avoid operating the fiber at the maximum of CEP fluctuations.
Using e.g. $\sim\!1\,$mJ input pulse energy and DPLP, we can achieve excellent long term CEP stability after the fiber, as shown in Figure 4. The residual CEP fluctuations have a standard deviation of $206\,$mrad, again confirming that the CEP stability of differentially pumped fibers is sufficient for attosecond experiments. Indeed, the full characterization of these pulses in Figure $4\,$(c) was achieved through attosecond streaking, which requires excellent CEP stability. Attosecond streaking using the few-cycle infrared (IR) pulses and isolated attosecond extreme ultraviolet (XUV) pulses was performed in a neon gas jet, using the setup described in detail in \cite{frank_technology_2012}. The $0.4\,$mJ compressed pulses have a duration of $3.5\,$fs, and were fully characterized using the vector potential retrieved from the FROG-CRAB trace, as described in \cite{witting_sub_2012}.

In conclusion, we have demonstrated that the output pulse energy from a $260\,$\textmu m inner-diameter hollow-fiber saturates at $0.6\,$mJ with the fiber statically filled with neon, and that the output pulse energy can be increased to $0.8\,$mJ by using either differential pumping, or CP pulses. We observe an overall degradation in the CEP stability as the input pulse energy is increased, which is the consequence of increased ionization within the fiber. For our experimental parameters, ionization losses prevent further degradation of the CEP stability above $1.25\,$mJ input pulse energies. However, this instability should be carefully considered when scaling hollow-fiber pulse compression to multi-millijoule energies. If the peak intensity is held constant by increasing the fiber inner diameter, the highly-nonlinear scaling of tunnel ionization with intensity can be avoided. Finally, we have presented the first direct evidence that the CEP stability performance of differentially pumped fibers can be equivalent to that of statically filled fibers. We have generated $0.4\,$mJ, $3.5\,$fs pulses with a CEP stability of $\sim\!200\,$mrad over $>2\,$h using a differentially pumped fiber, showing that the long term CEP stability of differentially pumped fibers is sufficient for attosecond experiments. We expect this work to aid the design of multi-millijoule hollow fiber systems where CEP stability is required.

\section*{Acknowledgements}

This work was financially supported by EPSRC through grants EP/I032517/1 and EP/F034601/1, and by Laserlab Europe (project LOA001657).
We acknowledge technical support from Andrew Gregory and Peter Ruthven.


\begin{thebibliography}{99}
%% Do not include separate BibTeX files; if BibTeX is used,
%% paste the output (contents of .bbl file) here.
\bibitem{nisoli_generation_1996}
M. Nisoli, S. De Silvestri, and O. Svelto, {Appl. Phys. Lett.} {{\bf 68}, 2793 (1996)}.
\bibitem{witting_characterization_2011}
T. Witting, F. Frank, C. A. Arrell, W. A. Okell, J. P. Marangos, and J. W. G. Tisch, {Opt. Lett.} {{\bf 36}, 1680 (2011)}.
\bibitem{bohman_generation_2010}
S. Bohman, A. Suda, T. Kanai, S. Yamaguchi, and K. Midorikawa, {Opt. Lett.} {{\bf 35}, 1887 (2010)}.
\bibitem{suda_generation_2005}
A. Suda, M. Hatayama, K. Nagasaka, and K. Midorikawa, {Appl. Phys. Lett.} {{\bf 86}, 111116 (2005)}.
\bibitem{robinson_the_2006}
J. S. Robinson, C. A. Haworth, H. Teng, R. A. Smith, J. P. Marangos, and J. W. G. Tisch, {Appl. Phys. B} {{\bf 85}, 525 (2006)}.
\bibitem{chen_generation_2009}
X. Chen, A. Jullien, A. Malvache, L. Canova, A. Borot, A. Trisorio, C. G. Durfee, and R. Lopez-Martens, {Opt. Lett.} {{\bf 34}, 1588 (2009)}.
\bibitem{borot_attosecond_2012}
A. Borot, A. Malvache, X. Chen, A. Jullien, J.-P. Geindre, P. Audebert, G. Mourou, F. Qu\'er\'e and R. Lopez-Martens, {Nature Phys.} {{\bf 8}, 416 (2012)}.
\bibitem{nagy_optimal_2011}
T. Nagy, V. Pervak, and P. Simon, {Opt. Lett.} {{\bf 36}, 3573 (2011)}
\bibitem{schweinberger_waveform_2012}
W. Schweinberger, A. Sommer, E. Bothschafter, J. Li, F. Krausz, R. Kienberger, and M. Schultze, {Opt. Lett.} {{\bf 37}, 3573 (2012)}.
\bibitem{wang_coupling_2009}
H. Wang, M. Chini, E. Moon, H. Mashiko, C. Li, and Z. Chang, {Opt. Express} {{\bf 17}, 12082 (2009)}.
\bibitem{vozzi_optimal_2005}
C. Vozzi, M. Nisoli, G. Sansone, S. Stagira, S. De Silvestri, {Appl. Phys. B} {{\bf 80}, 285 (2005)}.
\bibitem{nisoli_toward_1998}
M. Nisoli, S. Stagira, S. De Silvestri, O. Svelto, S. Sartania, Z. Cheng, G. Tempea, C. Spielmann, and F. Krausz, {IEEE J. Sel. Top. Quant. Electron.} {{\bf 4}, 414 (1998)}.
\bibitem{ammosov_tunnel_1986}
M. V. Ammosov, N. B. Delone, and V. P. Kra\u\i nov, {Sov. Phys. JETP} {{\bf 64}, 1191 (1986)}.
\bibitem{frank_technology_2012}
F. Frank, C. Arrell, T. Witting, W. A. Okell, J. McKenna, J. S. Robinson, C. A. Haworth, D. Austin, H. Teng, I. A. Walmsley, J. P. Marangos, and J. W. G. Tisch, {Rev. Sci. Instrum.} {{\bf 83}, 071101 (2012)}.
\bibitem{witting_sub_2012}
T. Witting, F. Frank, W. A. Okell, C. A. Arrell, J. P. Marangos, and J. W. G. Tisch, {J. Phys. B} {{\bf 45}, 074014 (2012)} 
\end{thebibliography}
\end{document}